# Full-bandwidth, continuous, and grayscale 3D nanolithography via line-illumination temporal focusing of ultrafast lasers

Qiuyuan Zhong[1,2,3], Charudatta Datar[1,3], Wei Liu[1,2], Gan Liu[1], Xiangsen Guo[1,2], Xuhao Fan[1,2], Fei Han[1], Bingxu Chen[1], Songyun Gu[1*], and Shih-Chi Chen[1,2*]

[1]Department of Mechanical and Automation Engineering, The Chinese University of Hong Kong, Shatin, N.T., Hong Kong SAR, China.

[2]Centre for Perceptual and Interactive Intelligence, Hong Kong Science Park, Shatin, N.T., Hong Kong

[3]These authors contributed equally.

*Correspondence to: scchen@cuhk.edu.hk; sygu1997@link.cuhk.edu.hk

**Abstract**

Achieving fast and continuous fabrication of large-scale complex 3D structures is key to unlocking industrial-scale adoption of two-photon lithography (TPL). Despite substantial improvement in peak optical patterning rates enabled by recent parallel exposure strategies, the practical fabrication rate of TPL for large structures remains low. This gap is primarily attributed to the mismatched bandwidth among toolpath generation, data transferring, and laser patterning, and the stop-and-go operation for part stitching etc. Here, we present a line-illumination temporal focusing TPL (Line-TF TPL) solution that, for the first time, demonstrates true continuous 3D nanolithography with full-bandwidth data streaming, grayscale voxel tuning, and cost-effective large-scale fabrication capability. To achieve the goal, we use a digital micromirror device (DMD) to temporally focus femtosecond laser pulses into a programmable line with enhanced 3D resolution, pixel-level grayscale control, and a high-refresh rate (>10 kHz), realizing continuous fabrication at a hardware-limited maximum rate. Specifically, we fabricated centimeter-scale 3D structures with sub-diffraction features down to 75 nm laterally and 99 nm axially. Our method eliminates stitching defects by continuous scanning and grayscale stitching; and provides real-time pattern streaming at a bandwidth that is one order of magnitude higher than previous TPL systems. The line-scanning strategy also substantially lowers the pulse-energy requirement, hence the cost for parallel TPL; and maximizes the machine uptime through continuous operation; both of which are critical metrics for industrialization. Finally, we demonstrated centimeter-scale artworks, fine 3D features, and complex miniaturized optics, revealing the Line-TF TPL's large-scale application potential in photonic packaging[1], metamaterial discovery[2,3], and biomedicine[4].

**Introduction**

The industrial adoption of a 3D nanofabrication method depends on its capability to reliably produce large-scale functional structures. Over the past two decades, two-photon lithography (TPL) has emerged as the leading technology driving advances in 3D nanofabrication. Leveraging the nonlinear two-photon absorption process, TPL enables the direct writing of arbitrarily complex polymeric architectures with sub-micrometer resolution[5], fundamentally extending the capabilities of traditional 2D photolithography. This technique has enabled a broad spectrum of innovations in micro-optics[1], micro-electronics[6], metamaterials[2,3], biomedicine[4], and other areas requiring complex miniaturized structures.

However, the use of TPL has been mostly limited to research laboratories; it has not yet achieved broad industrial adoption owing to persistent challenges in speed, quality, and reliability.

Conventional TPL relies on serially scanning a tightly focused femtosecond laser spot in photo-resins, resulting in a voxel-by-voxel fabrication rate below ~$10^6$ voxels/sec (<0.1 mm³/hr at ~200 nm lateral resolution and ~1 μm axial resolution). This inherently slow sequential process makes it impractical for commercial-scale production. Parallelization has therefore become a major research direction for improving the throughput. For example, projection-based TPL can achieve a peak projection speed of 333M voxel/s[7-9], and multi-focus TPL has demonstrated over 100M voxels/s[10-12]. However, most existing parallel TPL approaches only improve the laser exposure rate in a static field of view (FOV) of the objective lens. Therefore, in order to fabricate largescale parts (or parts of larger sizes than the FOV), a discontinuous stop-and-go operation with repetitive stage translations is required, which not only introduces vibration and mechanical artifacts, but also substantial machine idle time, sharply reducing the effective throughput even when the peak optical patterning rate is high. In addition, serious stitching defects are unavoidable in this sequential fabrication process, presenting a 2D defect grid in the final parts. Although some hybrid laser and substrate scanning methods have attempted to solve the problem of discontinuous fabrication and stitching defects[13], the path planning strategies for these methods are only feasible for single focal spot scanning and cannot be parallelized for high-throughput fabrication.

In some cases, the effort of parallelizing TPL introduces tradeoffs to its fabrication capability and technology accessibility. For example, diffractive optical elements (DOE) can be used generate fixed focal spot arrays for parallel writing, but their periodicity and lack of grayscale control compromise TPL's ability to produce arbitrary 3D geometries[10,14]. Note that grayscale manufacturing capability has become essential for the future development of TPL due to the growing demand for fabricating micro-optics and devices that require high surface quality or gradient material properties. Other approaches such as femtosecond light-sheet exposure[7-9], holographic multi-focus writing[11,15,16], and lens-array-based TPL[12,17], require large femtosecond laser amplifier systems with millijoule-level pulse energies that cost over $250,000 per laser. These hardware demands substantially raise the barrier for wide adoption.

Most critically, existing TPL platforms are increasingly constrained by the data-transfer rate from the computer to the fabrication volume, which set the "practical fabrication throughput". Some systems are limited by the response speed of the scanning mechanisms, such as galvanometric mirrors and acousto-optic deflectors (AOD)[18]; some are limited by the frame rate or bit-depth of spatial light modulators, such as the DMDs or liquid-crystal SLMs; and many others are limited by the non-continuous fabrication process inherent to sequential FOV stitching. With the rapid development of faster optical patterning methods in recent years, the practical fabrication throughput does not increase proportionately; instead, it remains well below the theoretical limit due to the combined result yielding from the in-FOV patterning rate and the inter-FOV translation speed. A scalable, cost-effective, and high-throughput TPL platform that can operate continuously at the full hardware bandwidth while retaining grayscale control for high-resolution 3D fabrication remains a critical unmet need for next-generation nanotechnologies.

**Line-illumination TF TPL**

Here, we present a line-illumination TPL solution based on the simultaneous spatial and temporal focusing of femtosecond pulses to address the aforementioned challenges. The Line-TF TPL not only has a unique parallel grayscale fabrication capability but also intrinsically overcomes the throughput limit by addressing the bottleneck of discontinuous printing, eliminating stitching defects. As depicted in **Fig. 1** and **Extended Data Fig. 1**, a femtosecond laser is incident onto a DMD (V9002, ViALUX, PCIe interface, optical patterning bandwidth 6.19 GB/s), where it is first patterned by the DMD mask, then dispersed in the *x-z* plane for temporal defocusing and remains unmodulated in the *y-z* plane. After the DMD, we inserted a negative cylindrical lens to diverge the beam in the *y-z* plane, generating a virtually focused line at the DMD plane. The dispersed and divergent beam then passes through a 4-f system composed of a collimating tube lens and an objective lens, creating a focused line pattern at the objective's focal plane. The line pattern is spatially focused in the *y-z* plane, which is in conjugation to the virtually focused line at the DMD, and it is temporally focused in the *x-z* plane, as different spectral components are dispersed by the DMD and recombined after the objective lens. The line can be programmed with arbitrary patterns and grayscale intensities by controlling the number of pixels turned on in the *y*-axis with a refresh rate of 13 kHz. Note that some DMDs may have a higher refresh rate of >20 kHz at the cost of reduced total pixels; the DMD used in this work presents the highest optical patterning bandwidth and fastest data transfer interface to date via commercially available components. However, even for this DMD, the data transfer rate (2.33 GB/s) is still much slower than the optical patterning bandwidth.

To address the issue, we developed a protocol for data compression and streaming to expand the effective data transfer rate from the sliced CAD model to the DMD controller, as illustrated in **Extended Data Fig. 2**. After slicing the part into layer patterns, we first find the unique line patterns within the scan area to effectively reduce the total DMD patterns to be used. Next, we generate the unique DMD patterns based on these line patterns and stream them to the DMD during TPL fabrication. With the current ratio between the data transfer rate and the optical bandwidth (2.33 GB/s vs. 6.19 GB/s), a compression factor ($\eta$) >2.66 would yield full bandwidth operation for the Line-TF TPL system. At lower compression factors, the effective fabrication rate becomes the data transfer rate multiplied by the compression factor (2.33×$\eta$ GB/s). (Detailed discussions about the compression and data streaming method are presented in the **Methods** section.) When the DMD patterns are continuously streamed from the computer to the DMD, the fabrication substrate is carried by a motion stage to perform linear scanning across centimeter areas. The scanning substrate is synchronized to the patterns displayed on the DMD to ensure high precision alignment between sequential scans and layers. This full-bandwidth, continuous operation eliminates the stop-and-go cycles of conventional TPL stitching process, driving the Line-TF TPL system to the throughput ceiling from commercially available hardware, which is 12.7 times higher than all previously reported TPL solutions for printing large parts (see **Extended Data Table 1**).

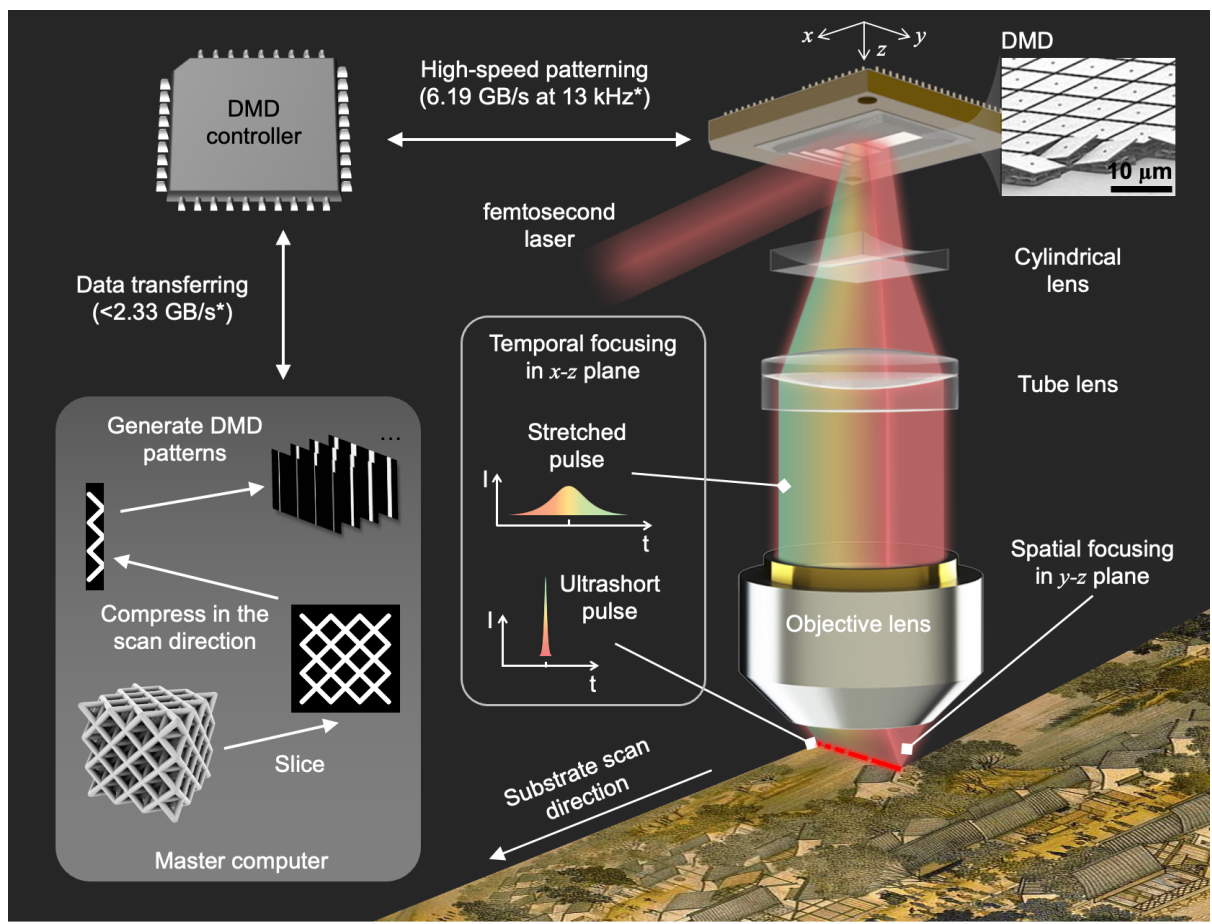

**Fig.1 | Full bandwidth operation of Line-TF TPL.** In the master computer, the CAD model is sliced into layered patterns and compressed in the scan direction, and a series of unique DMD patterns is

generated. The DMD patterns are streamed into the DMD controller in real-time during TPL printing at full optical patterning rate. *The data transfer rate and the optical patterning rate are from DMD model V9002 (ViALUX).* In the optical path, a femtosecond laser is incident onto the DMD, dispersed in the *x-z* plane into broadened pulses and remain unmodulated in the *y-z* plane. The laser then passes through a negative cylindrical lens to form a virtually focused line at the DMD. After that, the laser passes through a 4-f system (i.e., a collimating tube lens and an objective lens) to form a 3D resolved intensity-programmable line for TPL fabrication. The line is temporally focused on the *x-z* plane with ultrashort pulse widths and is spatially focused on the *y-z* plane. By scanning the sample substrate while modulating the DMD patterns, high precision grayscale structures can be rapidly fabricated.

Line-TF TPL's unique optical configuration further enables precise grayscale tuning that cannot be achieved by previous TF TPL systems (including other TF line-scanning systems), as well as other DMD-based parallel TPL methods, due to the binary modulation nature. Although holographic multi-focus TPL systems can achieve grayscale control in principle, it is done at the expense of high computation overhead of the iterative phase retrieval algorithms. In our system, the pixel-level grayscale intensity of the line pattern can be dynamically controlled by the number of DMD pixels turned on/off along the *y*-axis (**Fig. 2a**). By fine tuning the grayscale intensity during continuous scanning, we print bulky pillar structures and sub-diffraction suspended nanowires within the same task. Along the scan direction (*y*-axis), we measured 84 nm (lateral) and 99 nm (axial) linewidths for the suspended nanowires; and perpendicular to the scan direction (*x*-axis), we measured 75 nm (lateral) and 106 nm (axial) linewidths, as shown in **Fig. 2b.** The grayscale fabrication of fine features is highly repeatable and shows a low degree of linewidth variance in both directions, which validates Line-TF TPL's capability for high-throughput, robust nano-manufacturing. It is worth noting that conventional TPL systems usually exhibit worse axial resolution than lateral resolution, which leads to non-homogeneous properties in the lateral and the axial dimensions. However, in Line-TF TPL, the line pattern is both spatially and temporally focused, leading to enhanced axial resolution compared to simple spatial focusing (point-scanning) or projection-based temporal focusing (light-sheet) solutions. We simulated the point spread function (PSF) of the Line-TF TPL under an objective lens with a numerical aperture (NA) of 1.4, and confirm that the optical resolution is comparable in all three axes at near 400 nm (**Fig. 2c** and **Methods**), consistent with our experimental results.

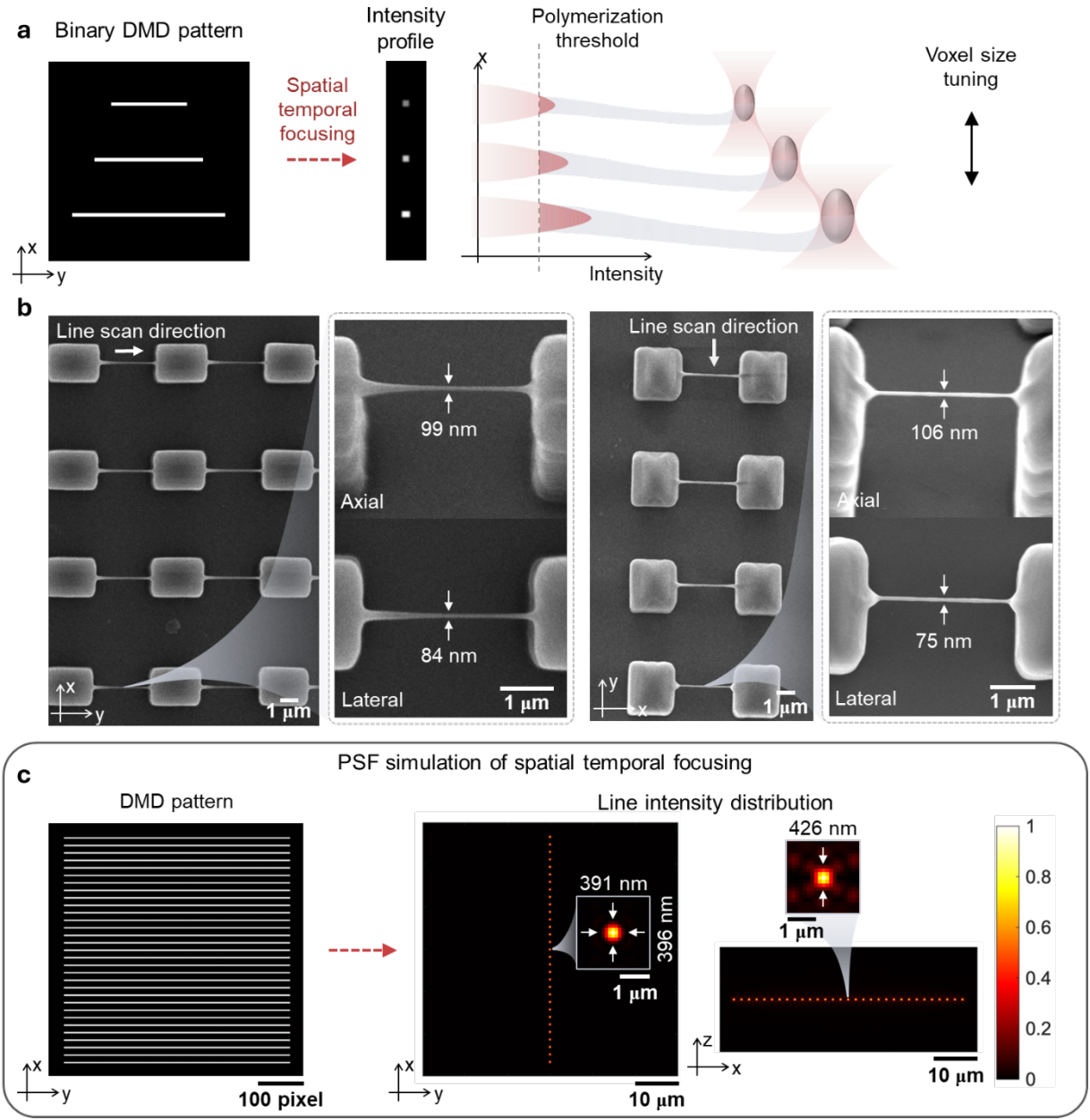

**Fig. 2 | Sub-diffraction fabrication enabled by grayscale tunability. a**, Grayscale voxel size tuning by changing the number of "on" pixels at the binary DMD. **b**, Measured lateral/axial linewidths are 84/99 nm along the substrate-scanning direction and 75/106 nm across it. Axial linewidths were measured from the SEM images taken at a 30° tilt angle. **c**, PSF simulation of temporal focusing. Time-averaged light intensity distribution in the *x*–*y* and *y*–*z* planes are calculated according to the projected pattern. Zoom-in images show the full width at half maximum (FWHM) calculated along the *x*-, *y*-, and *z*-directions.

**Continuous fabrication of miniaturized artworks**

In this section, we demonstrate the high-quality and high-speed characteristics of the Line-TF TPL by fabricating a series of miniaturized artworks featuring fine details and large footprints. Continuous fabrication was performed by synchronizing the scanning stage (*y*-axis) with the switching DMD patterns. To maximize the effective data transfer rate from the computer to the printing volume, we
5

first created a sequence of unique DMD patterns (**Extended Data Fig. 2**) from the sliced part and another sequence of indexes to address the DMD patterns during printing. Then, the DMD pattern sequence and the index sequence were streamed to the DMD controller and exposed on the fly, as detailed in the **Methods** section, where the pattern switching was triggered by the stage motion. As described in the last section, the system can operate at the full DMD bandwidth when the compression factor $\eta$ reaches 2.66. Here, we listed the compression factors for different objects (simplified in 2D form), including the ones demonstrated in this work, in **Extended Data Fig. 3**, where we can find that most of the tasks can operate at full bandwidth. We also segment the DMD memory into different zones, including a "projecting" zone, a "ready to project" zone, and an "uploading" zone (**Extended Data Fig. 2**) that operate separately to ensure continuous pattern streaming.

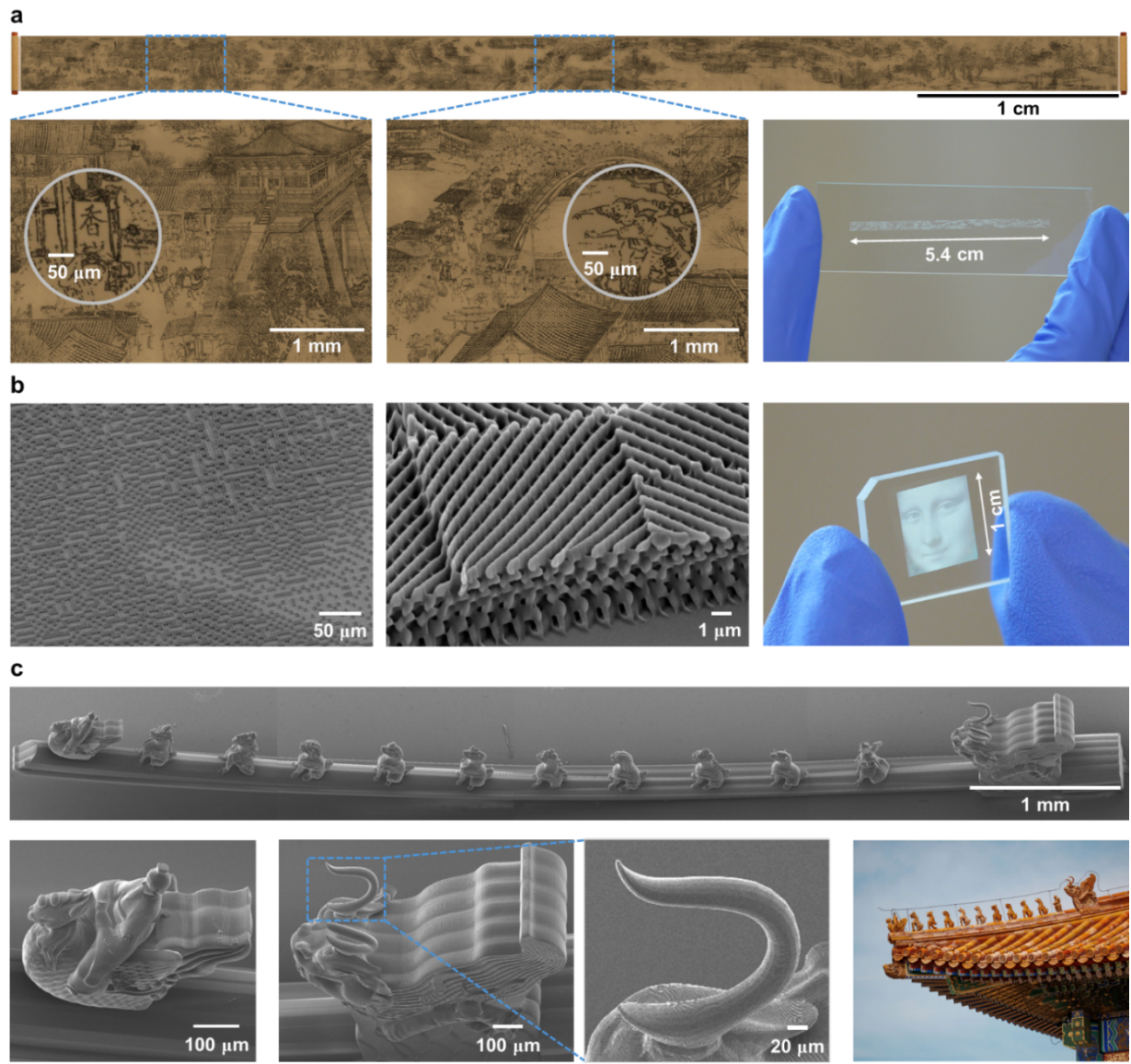

**Fig. 3 | Continuous fabrication of miniaturized artworks. a,** Photo and optical microscope images of a 5.4-cm replica of *Along the River During Qingming Festival (Song Dynasty, AD 960 - 1279)* fabricated via the Line-TF TPL, demonstrating sub-micron resolution across cm-scale areas. Zoomed-in views show intricate details (Chinese characters, figures, and trees) with no visual stitching defects. The exhibited images are pseudo color mapped to look similar to the original artwork. **b,** Photo and SEM images of a centimeter-scale monochrome *Mona Lisa* rendered by the dithering distribution of 3D woodpile metastructures with sub-micron linewidth. **c,** Design and SEM images of the *Ridge Beasts of the Hall of Supreme Harmony* (Forbidden City) demonstrating high-fidelity 3D fabrication.

With the Line-TF TPL, we eliminate the stop-and-go process, thereby overcoming the speed limitation of previously reported parallel printing methods. This result set a new TPL throughput record under continuous and grayscale operation (i.e., $3.3 \times 10^7$ voxels/s = 12,987 Hz × 2,560 voxels/line; each voxel has 1,600 grayscale levels), in which the fabrication rate is bounded by the fundamental device bandwidth limit rather than the limitations of parallel exposure processes. Our solution unlocks large-scale, stitch-free fabrication of completely non-repeating complex structures. **Fig. 3a** shows a 5.4-cm length replica of *Along the River During the Qingming Festival* (a Chinese masterpiece from the *Song Dynasty, AD 960 - 1279*), fabricated in 3 hours, demonstrating consistent sub-micron resolution across centimeter-scale areas. Zoom-in views reveal intricate details, including Chinese characters, human figures, trees, and animals, which exhibit no visual stitching artifacts. Note that the stitching defects in the scan direction (*y*) are eliminated by continuous scanning, while the stitching defects in the perpendicular direction (*x*) are minimized by the precise stage position feedback and grayscale stitching (details illustrated in **Fig. 4**). For such continuous, large-scale fabrication, we track the substrate interface in real-time and adjust for the tilt and unevenness of the substrate. **Fig. 3b** demonstrates a centimeter-scale monochrome grayscale image of the *Mona Lisa*, fabricated in 8 hours, where the gray levels are encoded by the dithering distribution of nano-architected metamaterials. The smooth tonal reproduction of the iconic portrait demonstrates precise sub-micrometer control over 3D architecture, enabling macroscopic monochrome display. **Fig. 3c** shows a 1-cm *Ridge Beasts of the Hall of Supreme Harmony in the Forbidden City*, which is fabricated in 13 hours. Each ridge beast exhibits a unique morphology, clear details, and smooth surface, such as the dragon's horns, demonstrating the system's capability to rapidly fabricate non-periodic complex structures. These results establish Line-TF TPL as a versatile platform for both 2D and 3D miniaturized artworks and similar structures.

**Fabrication of micro-optics with grayscale control**

Fine-tuning the laser intensity on the fly during printing has been a long-desired capability for high-precision TPL manufacturing. While many parallel TPL methods fail to achieve capability due to the limited optical patterning strategy, the Line-TF TPL allows straightforward and rapid grayscale intensity tuning for each pixel along the focused line pattern. Therefore, it is optimized for printing micro-optical structures with high uniformity, sub-diffraction features, and minimized stitching defects. By turning on/off certain DMD pixels perpendicular to the line pattern, we can precisely modulate the intensity pattern (**Fig. 4a** and **Extended Data Fig. 4**) for high printing quality. After uniformity correction, the coefficient of variation (*CV*, Standard Deviation / Mean × 100%, point number $n = 31$) of the line intensity can be improved from 21.93% (near gaussian laser profile) to 1.26% (uniform). In addition, to mitigate stitching artifacts between adjacent scan strips, we have a grayscale stitching zone between two adjacent scan strips by gradually reducing the active DMD pixel number near the strip edges (**Fig. 4a - 4b**). This creates a smooth intensity gradient in the overlap regions and enables seamless merging between the scan strips. When combined with continuous scanning in each strip, our method achieves seamless fabrication in both the *x* and *y* direction that is ideal to fabricate large-scale optical surfaces.

A continuous holographic structure is fabricated on a glass substrate within 13 minutes with active grayscale control (**Fig. 4c - 4d**). The resulting film measures 5 cm in length and contains 100 individual holographic frames, each composed of 500 × 500 pixels with a 1-μm pixel size. Under red laser illumination, each frame generates a corresponding diffraction pattern. By translating the sample with a motion stage, sequential holographic frames are successively displayed, producing a high-fidelity holographic display sequence (**Fig. 4e**). In contrast, without gradient overlapping at the boundaries between the scan strips, the holographic display result of the fabricated structure shows undesired artifacts (**Fig. 4f**). This rapid and scalable fabrication method demonstrates its application potential in information encoding and holographic display. Such continuous holographic films can also be printed on flexible polyethylene terephthalate (PET) substrates (insert in **Fig. 4c**), can encode complex 3D information when integrated with roll-to-roll manufacturing, and can be used for large-scale production of 3D holographic films.

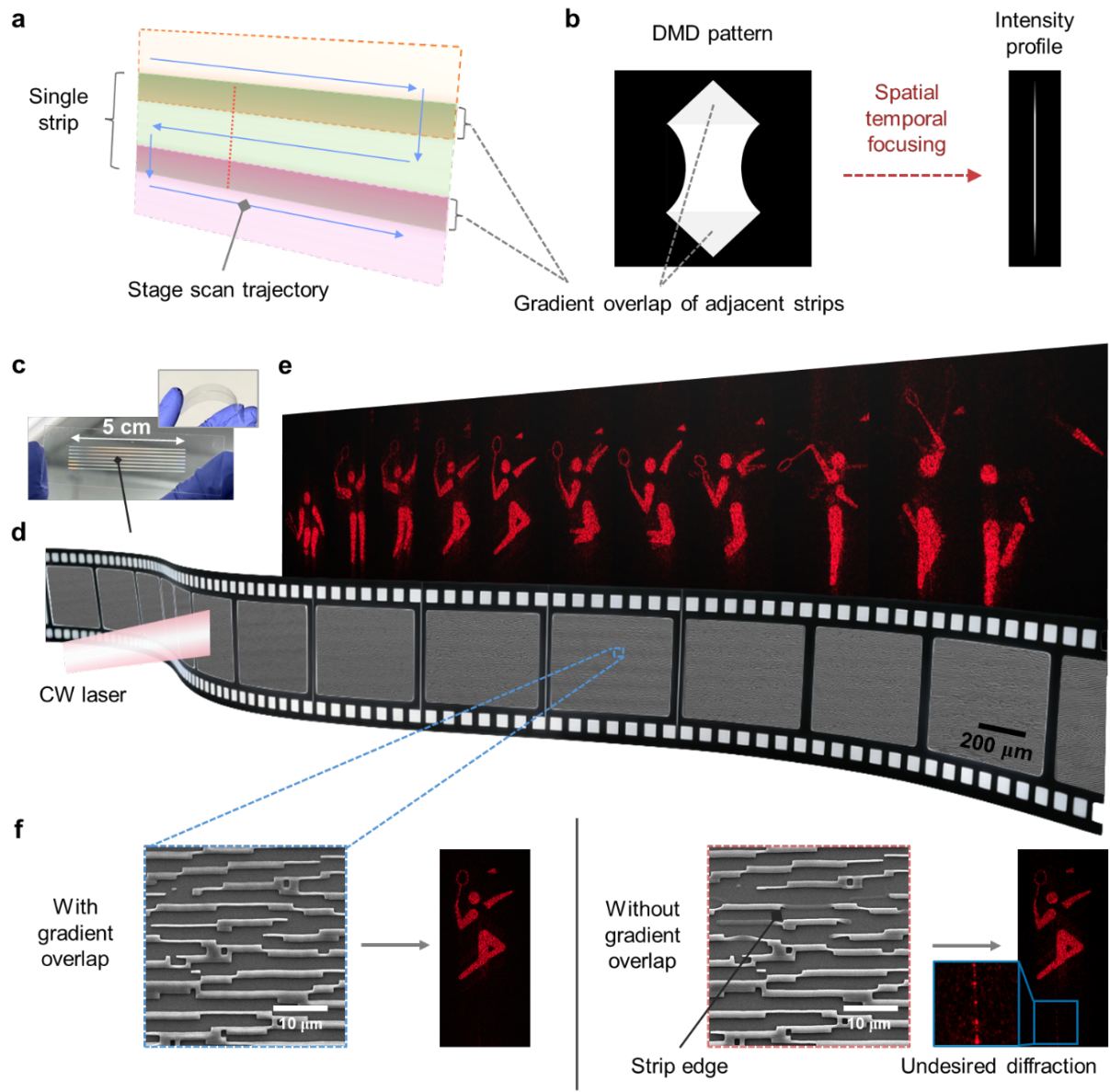

**Fig. 4 | High-fidelity grayscale fabrication of holographic films. a,** Schematic of the gradient overlapping strategy applied between adjacent scan strips. This method introduces a gradual reduction in exposure dose at the strip edges, minimizing stitching artifacts perpendicular to the scanning direction. **b,** DMD pattern (left) and the simulated intensity profile at the focal plane (right) showing the principle of grayscale control. The modulation ensures uniform intensity along the strip center while creating a smooth gradient at the two edges. **c,** Photograph of a holographic film printed on a glass substrate. Insert is a holographic film printed on a flexible PET substrate. **d,** Images of representative holographic frames. **e,** Experimentally captured diffraction patterns from holographic frames under red laser illumination showing the dynamic display sequence achieved by translating the film. **f,** Comparison of holographic display fidelity with (left) and without (right) gradient overlaps at strip boundaries, demonstrating effective suppression of undesired structure stitching artifacts.

**Discussion and outlook**

The Line-TF TPL is a high-throughput parallel 3D nanofabrication solution with unprecedented quality and reliability. By spatially and temporally focusing the femtosecond laser pulses into programmable line patterns, the Line-TF TPL achieves a near isotropic voxel shape, which is ideal for 3D manufacturing and is proven by the sub-100 nm linewidth in both lateral and axial dimensions. The system can deliver grayscale toolpath data to the fabrication volume with nearly full optical patterning bandwidth. The continuous scan and grayscale stitching strategy eliminates stitching artifacts over industrial scale areas (>5 cm). It further achieves a 12.7-fold increase in sustainable printing bandwidth ($2.33 \times \eta$ GB/s with an upper limit of 6.19 GB/s) compared to other existing parallel TPL methods (**Fig. 5** and **Extended Data Table 1**). This bandwidth corresponds to a continuous-operation voxel rate of up to $3.3 \times 10^7$ voxels/s, where each voxel has 1,600 grayscale levels that leads to adjustable 3D resolution (100 nm – 400 nm), enabling a sustainable volume printing rate of 0.11 – 7.6 mm³/hr. Note that by changing the objective lens to lower magnification rate and NA, much higher volume printing rate can also be achieved; for example, a 10× objective lens with 2-μm resolution would lead to a fabrication rate of 950 mm³/hr). Another distinctive advantage of the grayscale programming capability is that it can be used to compensate for suboptimal quality of laser beams (as illustrated in **Extended Data Fig. 4**), as generating a uniform temporally focused light sheet or line is the foundation of generating quality prints; yet, maintaining a femtosecond laser amplifier's peak performance can sometimes be challenging. This feature will ensure the Line-TF TPL system can operate in a longer time frame with peak performance and without laser or system maintenance.

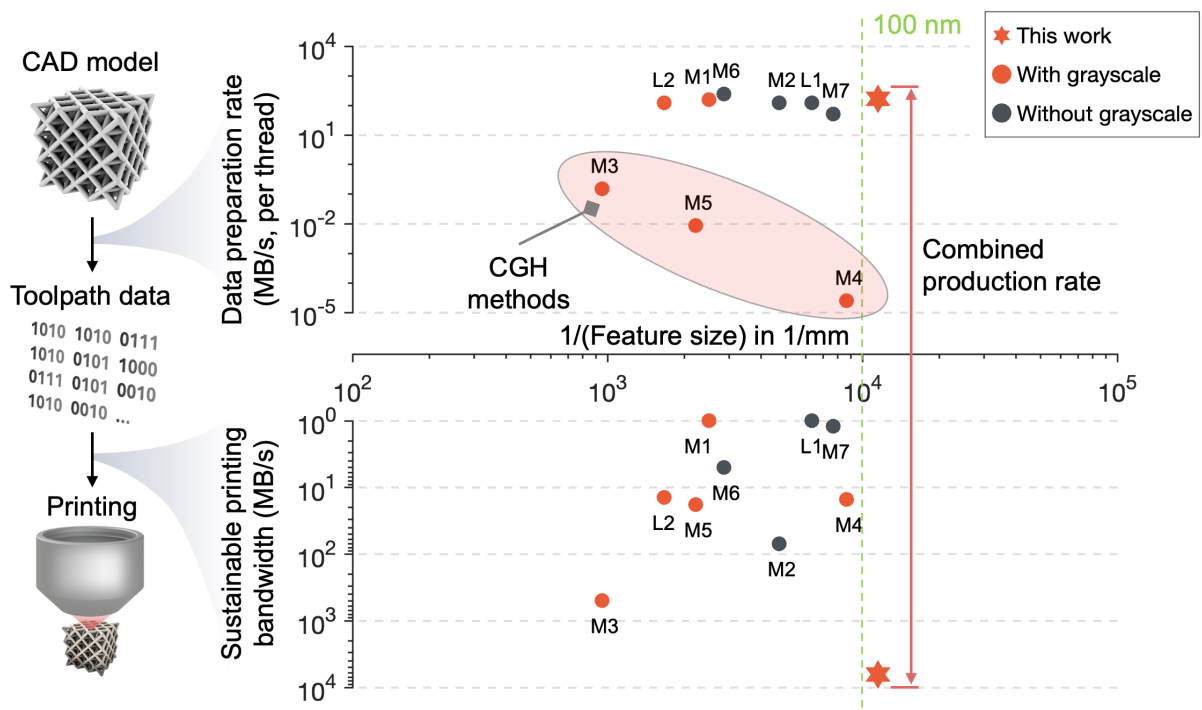

**Fig. 5 | Comparison of feature size (average of lateral and axial feature sizes), data preparation rate, and sustainable printing bandwidth for different TPL methods.** Study references: Line-TF TPL (this work, hexagram); projection-based TF: L1(2019 Saha et al.[7], 2022 Han et al.[9]), L2 (2021 Somers et al.[8]); Multi-focus: M1 (2023 Yang et al.[17]), M2 (2023 Jiao et al.[18]), M3 (2019 Geng et al.[15]), M4 (2024 Ouyang et al.[16]), M5 (2024 Zhang et al.[11]), M6 (2024 Kiefer et al.[10]), and M7 (2024 Wang et al.[14]).

Furthermore, this parallel TPL strategy exhibits high data processing efficiency that is compatible with the high printing throughput, which significantly improves the machine's up time. For the Line-TF TPL, the grayscale toolpath data can be prepared from a CAD model at a rate of 168 MB/s per central processing unit (CPU) thread. Utilizing multi-core parallel processing, the throughput can readily support the data demands of high-speed printing. In contrast, some parallel TPL methods show unmatched data preparation rate and printing rate, which would ultimately result in undesired machine idling. For example, TPL methods based on computer-generated holography (CGH) can only achieve less than 0.15 MB/s per CPU thread when preparing the toolpath data due to the iterative phase retrieval processes, as shown in **Fig. 5**. This aspect becomes increasingly important in the high-throughput printing regime as the complexity of the printing task will also scale up significantly. In practice, the production rate is deducted from both the data preparation rate and the sustainable printing rate (**Fig. 5**).

Lastly, the adoption of the Line-TF TPL would be significantly easier than other parallel TPL methods due to the lower pulse energy requirement, which allows the laser sources to be cost-effective femtosecond lasers with smaller footprints, rather than expensive femtosecond laser amplifiers. We foresee that in the near future, the Line-TF TPL would be directly combined with roll-to-roll systems for large-scale microstructure production. The low adoption barrier, high-quality, and rapid fabrication capability enables the Line-TF TPL to be widely used in many applications such as biomedicine, micro-photonics and electronics, and artificial material production.

## Acknowledgments

We thank members from the Chen's group at CUHK and staff members of Astra Optics Limited for assistance with experiments and comments on the manuscript. We are grateful to X. Fu, J. Li, Y. Zhang, Y. Wang, and C. Wang for experimental helps, and H. Li and H. Chen for constructive comments.

This work was supported by Research Grants Council, (RGC), Senior Research Fellow Scheme (SRFS2526-4S01), Collaborative Research Fund (C4074-22GF); General Research Fund (14211224); and the Centre for Perceptual and Interactive Intelligence (CPII), a CUHK-led InnoCentre under the InnoHK initiative of the Innovation and Technology Commission of the Hong Kong Special Administrative Region Government.

## Author contributions

S-C.C., Q.Z., S.G., and C. D. conceived the study and prepared the manuscript. Q.Z., S.G., W.L., G.L., B.C., X. F., and X. G. designed and constructed the Line-TF TPL setup and performed fabrication experiments. F.H. prepared the printing material.

## Ethics declarations

A US pattern application (No. 18/467,170, 14 Sept. 2023) was filed with S.G., Q.Z., and S-C.C. as co-inventors.

## References

1   Yu, S. *et al.* Two-photon lithography for integrated photonic packaging. *Light: Advanced Manufacturing* **4**, 486-502 (2024).

2   Xia, X. *et al.* Electrochemically reconfigurable architected materials. *Nature* **573**, 205-213 (2019).

3   Zhou, W. *et al.* 3D polycatenated architected materials. *Science* **387**, 269-277 (2025).

4   Urciuolo, A. *et al.* Intravital three-dimensional bioprinting. *Nature biomedical engineering* **4**, 901-915 (2020).

5   Kawata, S., Sun, H.-B., Tanaka, T. & Takada, K. Finer features for functional microdevices. *Nature* **412**, 697-698 (2001).

6   Xu, S. *et al.* 3D-printed micro ion trap technology for quantum information applications. *Nature*, 1-7 (2025).

7   Saha, S. K. *et al.* Scalable submicrometer additive manufacturing. *Science* **366**, 105-109 (2019).

8   Somers, P. *et al.* Rapid, continuous projection multi-photon 3D printing enabled by spatiotemporal focusing of femtosecond pulses. *Light: Science & Applications* **10**, 199 (2021).

9   Han, F. *et al.* Three-dimensional nanofabrication via ultrafast laser patterning and kinetically regulated material assembly. *Science* **378**, 1325-1331 (2022).

10  Kiefer, P. *et al.* A multi-photon (7× 7)-focus 3D laser printer based on a 3D-printed diffractive optical element and a 3D-printed multi-lens array. *Light: Advanced Manufacturing* **4**, 28-41 (2024).

11  Zhang, L. *et al.* High-throughput two-photon 3D printing enabled by holographic multi-foci high-speed scanning. *Nano Letters* **24**, 2671-2679 (2024).

12  Gu, S. *et al.* 3D nanolithography with metalens arrays and spatially adaptive illumination. *Nature* **648**, 591-599, doi:10.1038/s41586-025-09842-x (2025).

13  Jonušauskas, L., Baravykas, T., Andrijec, D., Gadišauskas, T. & Purlys, V. Stitchless support-free 3D printing of free-form micromechanical structures with feature size on-demand. *Scientific reports* **9**, 17533 (2019).

14  Wang, X. *et al.* 3D Nanolithography via Holographic Multi-Focus Metalens. *Laser & Photonics Reviews* **18**, 2400181 (2024).

15  Geng, Q., Wang, D., Chen, P. & Chen, S.-C. Ultrafast multi-focus 3-D nano-fabrication based on two-photon polymerization. *Nature communications* **10**, 2179 (2019).

16  Ouyang, W. *et al.* Ultrafast 3D nanofabrication via digital holography. *Nature communications* **14**, 1716 (2023).

17  Yang, S. *et al.* Parallel two-photon lithography achieving uniform sub-200 nm features with thousands of individually controlled foci. *Optics Express* **31**, 14174-14184 (2023).

18  Jiao, B. *et al.* Acousto-optic scanning spatial-switching multiphoton lithography. *International Journal of Extreme Manufacturing* **5**, 035008 (2023).

## Methods

### Materials

We used commercial IP-DIP (Nanoscribe, Germany) photo-resin for TPL printing. After printing, development was performed in propylene glycol monomethyl ether acetate (PEGMA) for 15 minutes, followed by rinsing in isopropyl alcohol (IPA) for 10 minutes.

### Line-TF TPL system

A femtosecond laser source (Chameleon Ultra II, 3.5 W at 800 nm wavelength, 80 MHz repetition rate, 140 fs pulse width, Coherent, USA) was used as the light source, which was then expanded by a 2×–8× beam expander (ZBE3B, Thorlabs, USA) to illuminate a DMD (V9002, resolution 2,560 × 1,600, maximum switching rate 12,987 Hz, pitch 7.6 μm, ViALUX, Germany). The DMD spatially modulated the beam intensity by switching the designed patterns. This modulated beam then passed through a concave cylindrical lens (f = -75 mm, Thorlabs, USA) positioned at its focal length, which transformed the beam into a virtual line profile at the plane of the DMD. The line-shaped beam was relayed through a 4-f imaging system comprising a collimator (f = 200 mm, Thorlabs, USA) and an oil-immersion objective lens (40×, NA 1.4, Zeiss, Germany). The substrate was mounted on a high-precision air-bearing motorized XYZ stage with a repeatability of ±50 nm (ABL1000 series, Aerotech, USA). The stage was synchronized via trigger signals to the DMD to enable continuous pattern switching during fabrication. Line-scanning was achieved by moving the substrate at a constant speed along the y-axis, and the 3D structure was achieved by layer-by-layer printing.

### Modeling of the light intensity distribution during line-scanned spatiotemporal focusing

The light intensity distribution during temporal focusing is obtained by simulating the propagation of a femtosecond laser pulse through the optical system. The optical system comprises a sequence of optical elements including a DMD, a cylindrical lens, a tube lens, and an objective lens. To simulate the intensity distribution at the sample plane, we first separately evaluate the electric field for each wavelength using monochromatic coherent optical models and then sum the contributions of all wavelengths to compute the total light intensity in the focal volume.

Light propagation through the optical system is modeled through the following steps:

  i. Femtosecond laser pulses are incident on the DMD chip surface.

 ii. The DMD functions as an amplitude mask that modulates the field amplitude according to the displayed pattern; the tilted micromirrors on the DMD simultaneously function as a blazed grating, introducing a wavelength-dependent angular spread, which results in horizontal spectral dispersion of the reflected pulses.

iii. The horizontally oriented cylindrical lens, a negative lens with its virtual focus coincident with the DMD plane, compresses the spectrally dispersed beam in the direction orthogonal to the dispersion. This yields a one-dimensional extended profile along the horizontal (dispersion) axis while narrowing the vertical profile.

 iv. The tube lens then recollimated the shaped beam, effectively relaying the spatial distribution toward the objective. The resulting collimated field encodes the one-dimensional line distribution prepared by the upstream elements.

  v. The objective lens focuses this one-dimensional profile onto the sample/substrate, generating a line projection at the objective's front focal plane.

In this system, the DMD spectrally separates the pulse while the combination of the cylindrical lens and tube lens preserves the one-dimensional profile and prepares it for focusing. The objective lens finally recombines the spectral components in the focal volume. Temporal focusing is achieved through this mechanism of spectral separation, cylindrical compression, and subsequent recombination.

We use scalar electric fields and paraxial approximations to develop a parametric model. Note that the optical system can also be modeled with vectorial optics to generate more accurate results for high numerical aperture objectives. Nevertheless, the paraxial approximation is sufficiently accurate to capture the key scaling effects of temporal focusing in the presence of cylindrical compression.

The electric field ($U_0$) of the laser beam that is incident on the DMD surface can be mathematically expressed as:

$$U_0(x_d, y_d, \omega) = A_0 \exp\left(-\frac{(\omega - \omega_0)^2}{\Omega^2}\right) \tag{1}$$

where $x_d$ and $y_d$ are spatial coordinates on the surface of the DMD; $\omega$ is the optical frequency; and $A_0$ is the peak amplitude.

The field emerging from the DMD ($U_d$) is expressed in Eq. (2):

$$U_d(x_d, y_d, \lambda) = U_0(x_d, y_d, \omega) H(x_d, y_d) \exp\left(\frac{2\pi i (x_d + y_d) \sin\theta_{m,\lambda}}{\lambda}\right) \tag{2}$$

where $\theta_{m,\lambda}$ is the angle of diffraction for the wavelength $\lambda$ and $H$ is the amplitude mask that is digitally encoded on the DMD. Note that the function $H$ has a binary value of 0 or 1 at each spatial coordinate on the DMD surface.

The phase component on the right-hand side represents the linear phase due to the grating-based dispersion from the DMD. The DMD plane, the incident beam, and the collimating lens are oriented such that the diffracted beam corresponding to the central wavelength ($\lambda_0$) emerges along the optical axis of the collimating lens. With such an orientation, the angle of diffraction can be mathematically expressed in Eq. (3):

$$\sin\theta_{m,\lambda} = \frac{m}{d}(\lambda - \lambda_0) \tag{3}$$

where $m$ is the diffraction order and $d$ is the DMD grating pitch. Thus, the field emerging from the DMD ($U_d$) can be expressed as:

$$U_d(x_d, y_d, \lambda) = U_0(x_d, y_d, \omega) H(x_d, y_d) \exp\left(2\pi i (x_d + y_d) \frac{(\lambda - \lambda_0)}{\lambda} \frac{m}{d}\right) \tag{4}$$

Note that light emerging from the DMD is subsequently modulated by a cylindrical lens whose cylindrical axis is oriented horizontally. Under the thin-lens and paraxial approximations, the cylindrical lens introduces optical power only along the vertical ($y$) direction, while leaving the horizontal ($x$) direction unaffected. For each wavelength component $\lambda$, the complex transmission function of the cylindrical lens can be expressed as:

$$H_{\text{cyl}}(y_d, \lambda) = \exp\left(-i\frac{2\pi(\lambda)}{\lambda f_{\text{cyl}}} y_d^2\right) \tag{5}$$

where $f_{\text{cyl}}$ denotes the focal length of the cylindrical lens. In our configuration, a negative cylindrical lens is employed, such that its virtual focal plane coincides with the DMD chip surface. Accordingly, the optical field immediately after the cylindrical lens can be expressed as:

$$U'_d(x_d, y_d, \lambda) = U_d(x_d, y_d, \lambda) \cdot \exp\left(-i\frac{2\pi(\lambda)}{\lambda f_{\text{cyl}}} y_d^2\right) \tag{6}$$

The light field is collected by the collimating lens and then focused by the objective lens. When the DMD chip surface is located at the focal plane of the collimating lens, the field at the back focal plane of the objective lens ($U_B$) is obtained through a Fourier transform of the field at the DMD, expressed as:

$$U_B(x_d, y_d, \lambda) = \frac{1}{i\lambda f_1} F\{U'_d(x_d, y_d, \lambda)\} \quad (7),$$

where $f_1$ is the focal length of the collimating lens; $i = \sqrt{-1}$ and $F$ is the Fourier transform operator. The spatial coordinates in the back focal plane of the objective ($x_b$, $y_b$) are related to the coordinates on the DMD as expressed in Eq. (8) and (9):

$$x_b = x_d \frac{\lambda}{f_1} \quad (8)$$

$$y_b = y_d \frac{\lambda}{f_1} \quad (9)$$

The field propagating into the back aperture of the objective lens ($U_A$) is limited by the pupil function ($P$) of the objective as expressed in Eq. (10):

$$U_A(x_b, y_b, \lambda) = U_B(x_b, y_b, \lambda) P(x_b, y_b) \quad (10)$$

The pupil function depends on the diameter of the circular back aperture of the objective lens ($D_{obj}$) as mathematically described by Eq. (11):

$$P(x_b, y_b) = \begin{cases} 1 & \text{for } |\sqrt{x_b^2 + y_b^2}| < 0.5 D_{obj} \\ 0.5 & \text{for } |\sqrt{x_b^2 + y_b^2}| = 0.5 D_{obj} \\ 0 & \text{for } |\sqrt{x_b^2 + y_b^2}| > 0.5 D_{obj} \end{cases} \quad (11)$$

The field in the focal plane of the objective lens ($U_f$) is obtained by propagating the field through a thin lens as expressed in Eq. (12):

$$U_f(x_f, y_f, z_f, \lambda) = \frac{n}{i\lambda f_2} \exp\left(\frac{2\pi i(2f_2 + z_f)}{\lambda/n}\right) F\left\{U_A(x_b, y_b, \lambda) \exp\left(-\frac{2\pi i(x_b^2 + y_b^2)}{\lambda f_2/n} \frac{z_f}{f_2}\right)\right\} \quad (12),$$

where $n$ is the refractive index of the immersion medium of the objective lens, $f_2$ is the focal length of the objective lens and $z_f$ is the axial coordinate along the propagation direction with its origin at the focal plane of the objective. At the focal plane (i.e., at $z = 0$), Eq. (12) reduces to a Fourier transform of the field at the back aperture of the objective ($U_A$) with a constant phase pre-factor. The lateral coordinates $x_f = Mx_d$ and $y_f = My_d$ are related to the spatial coordinates on the DMD chip surface through optical magnification of the system ($M = f_2/f_1$).

Finally, the temporal evolution of the pulse in the focal volume ($U_{TF}$) is obtained by combining the monochromatic fields, as expressed in Eq. (13):

$$U_{TF}(x_f, y_f, z_f, t) = \int_{-\infty}^{\infty} U_f(x_f, y_f, z_f, \omega) \exp(-2\pi i \omega_0 t) d\omega \quad (13)$$

The time-varying intensity in the focal volume ($I_{TF}$) is obtained from the electric field as expressed in Eq. (14):

$$I_{TF}\left(x_f, y_f, z_f, t\right) = \left|U_{TF}\left(x_f, y_f, z_f, t\right)\right|^2 \quad (14)$$

**Data compression and streaming**

The high-speed patterning capability of a DMD is constrained by two critical bottlenecks: the limited onboard memory of its controller and the bandwidth of the data transfer interface. Complex, large-scale patterns with nanoscale resolution can require hundreds of gigabytes to terabytes of data, and typical DMD controller memory capacities of 8 - 32 GB are insufficient for pre-loading entire datasets. Even for systems where the controller is equipped with solid-state drives (SSDs) to expand storage to several terabytes, the data transfer interface — often USB 3.0 or HDMI — remains a severe bottleneck. Although the intrinsic projection bandwidth of a DMD can exceed 6 GB/s, common interfaces such as USB 3.0 (~500 MB/s) and enhanced PCIe links (~2.33 GB/s) fall far short of this peak, ultimately limiting continuous high-speed patterning.

To overcome these limitations, we developed a data compression and streaming protocol that enables continuous fabrication. Our method eliminates the need to store the entire dataset in the controller's memory prior to printing. The workflow is as follows: First, a 3D model is sliced into individual layer patterns. For each layer, we identify and encode unique, non-repeating line patterns within whole scans, significantly reducing the total data volume that must be transferred.

During fabrication, these compressed patterns are streamed continuously from the host computer to the DMD controller. To manage the limited onboard memory, we partition the DMD memory into blocks. The system operates cyclically: while one block of pattern data is being projected, the next block is asynchronously uploaded into a separate memory section. Upon completion of a block's projection, its memory is released and subsequently overwritten with new pattern data in the next cycle. This "stream-and-swap" approach ensures a continuous data flow, decoupling the fabrication process from the overall pattern dataset size.

For complex patterns with minimal data redundancy, such as the detailed scene *Along the River During the Qingming Festival*, the achievable printing speed is constrained by the available data transfer bandwidth, as the compression factor approaches 1. In such cases, the effective fabrication rate is limited to the PCIe upload rate of 2.33 GB/s. Conversely, for structures with high periodicity or significant redundancy — e.g., a monochrome image of the *Mona Lisa* composed of 3D woodpile metastructures — a high compression factor ($\eta \gg 2.66$) can be achieved. This enables the system to operate at the DMD's maximum projection rate of 6.19 GB/s, fully leveraging its bandwidth without being limited by data transfer.

This integrated approach of data compression and real-time streaming fully leverages the DMD's high-speed potential, enabling the continuous fabrication of large-scale structures without interruption.

**Device synchronization for continuous printing**

For hardware-level synchronization between the continuous y-stage motion and DMD projection, the stage's position-synchronized output (PSO) signal was employed to trigger the DMD in real time. Specifically, as the stage translates the sample at a constant velocity along the y-axis, it generates a synchronized pulse at fixed spatial intervals (e.g., every 100 nm) to trigger the DMD to project the corresponding pattern. The DMD is configured in an uninterrupted binary projection mode, ensuring each trigger results in immediate pattern exposure without latency, thereby achieving its maximum refresh rate (13 kHz). This approach enables continuous, synchronous printing during stage motion, eliminating stop-and-go operation and thus enhancing throughput.

**Data availability**

The data that supports the findings of this study are available from the corresponding author on request.

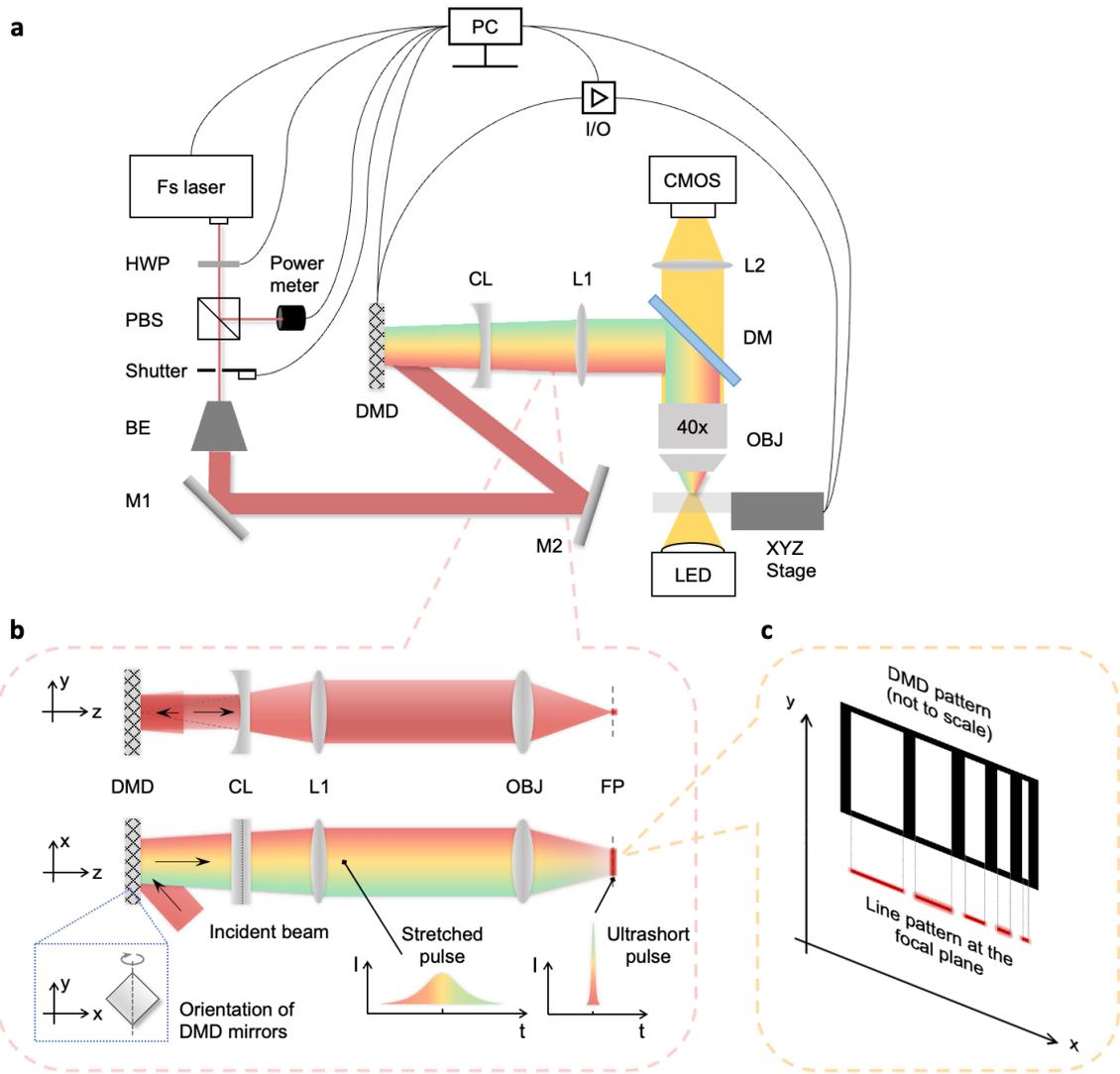

**Extended Data Fig. 1 | Optical setup of the Line-TF TPL system. a**, the overall design of the Line-TF TPL system; **b**, *y-z*, *x-z*, and *x-y* views of femtosecond laser propagation in the Line-TF TPL system. The rotational axis of the DMD mirrors is parallel to the *y*-axis, which induces dispersion in the *x-z* plane while keeping the beam non-modulated in the *y-z* plane. The femtosecond laser beam is stretched in the time domain after the DMD and is then temporally (i.e., *x-z* plane) and spatially (i.e., *y-z* plane) focused into an intensity programmable-line at the focal plane of the objective lens; **c**, spatially and temporally focused line pattern in the *x-y* plane and the corresponding DMD pattern. Fs laser: Femtosecond laser, HWP: half wave plate, PBS: Polarization beam splitter, BE: Beam expander, M1-M2: Mirrors. DMD: Digital micromirror device, CL: Cylindrical lens, L1-L2: lens, DM: Dichroic mirror, OBJ: Objective lens, LED: Light-emitting diode, CMOS: Complementary metal oxide semiconductor, I/O: Multifunction input/output device, PC: Computer, FP: Focal plane.

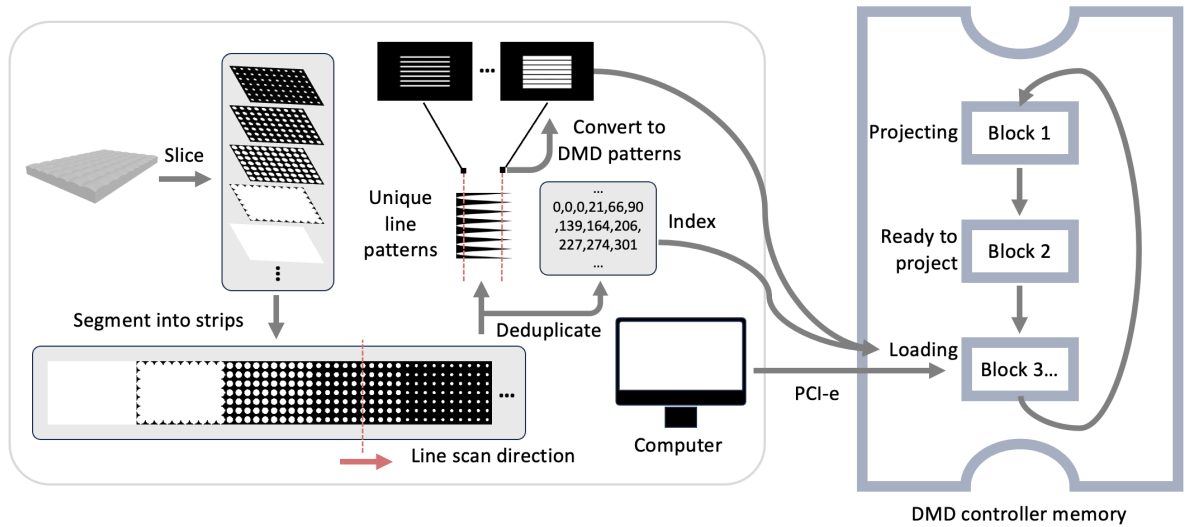

**Extended Data Fig. 2 | Workflow of the compression and streaming protocol.** A 3D CAD model is sliced into layers, and unique line patterns within whole scans are identified to create a minimal set with a dictionary. During fabrication, the DMD controller memory is a cyclic fashion: while one block of patterns is projected, the next is asynchronously uploaded, enabling continuous patterning beyond the DMD controller's memory limit.

| Model | Sample size (mm) | Strip width of single scan (μm) | Compression factor η |
|---|---|---|---|
| 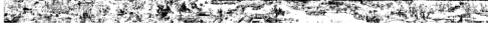 *Along the River During the Qingming Festival* | 54.42 × 2.59 | 150 | 1.06 |
| 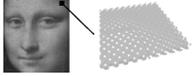 *Mona Lisa encoded by 3D woodpile metamaterials* | 10.08 × 8.96 | 40.5 | 63,934.27 |
| 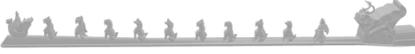 *Ridge Beast of the Hall of Supreme Harmony* | 8.39 × 0.52 × 0.92 | 150 | 107.29 |
| 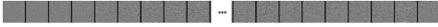 *Holographic film* | 49.13 × 0.50 | 40 | 1.25 |
| 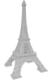 *Eiffel tower* | 0.55 × 0.55 × 1.21 | 100 | 373.58 |
| 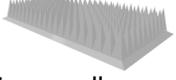 *Micro-needle array* | 2.12 × 1.12 × 0.32 | 100 | 23,858.08 |

**Extended Data Fig. 3 | Summarized compression factors for different objects.** Although lower factors are observed for painting and holography, most demonstrated tasks attained a compression factor *η* above 2.66, which is sufficient to leverage the full optical bandwidth of the system.

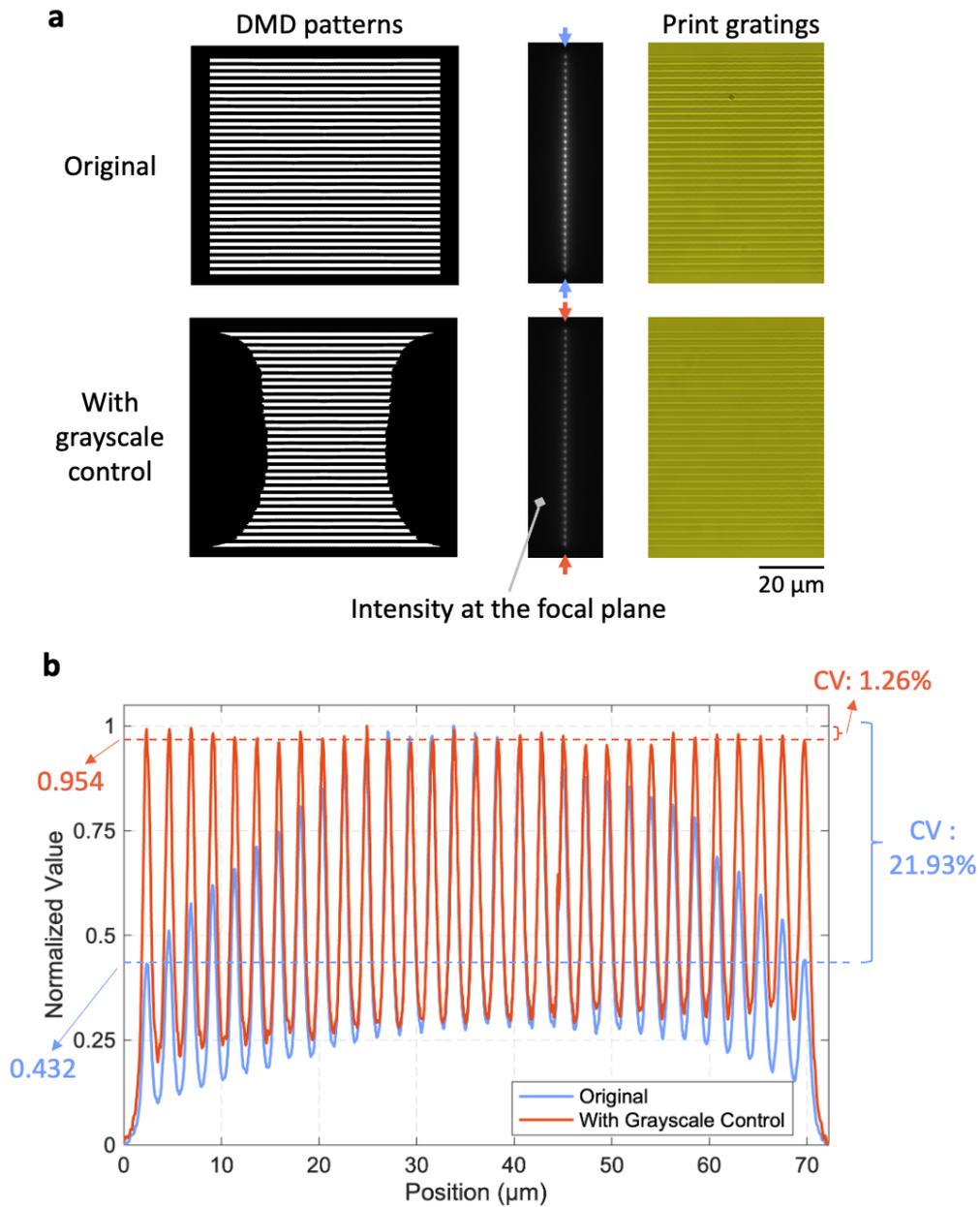

**Extended Data Fig. 4 | Pixel-level grayscale control to obtain uniform laser intensity along the line at the focal plane. a**, DMD patterns, corresponding intensity of the focal plane, and printing gratings with and without intensity pre-compensation for Gaussian-like beam. **b**, Normalized value of the intensity at the focal plane indicated by arrows in **a**.

**Extended Data Table 1 | Comparative analysis of parallel TPL technologies.**

| Study | Method | Lateral/axial feature size (nm) @ wavelength (nm) | Sustainable printing bandwidth (MB/s) *1 | Data preparing rate per CPU core (MB/s) *2 | Grayscale capability | Stitch-free | Pulse energy | Laser type & cost (USD) |
|---|---|---|---|---|---|---|---|---|
| **Line-TF TPL (This work)** | DMD line projection | **75/99 @ 780** | **2,393×$\eta$, upper limit: 6,341** | **168** | **Pixel-level control** | **XY axis** | **50 nJ** | Oscillator ~$120,000 |
| Light-sheet[7,9] | DMD area projection | 142/174 @ 800 | 1 | 123 | No | No | 4 mJ | Amplifier ~$250,000 |
| Light-sheet[8] | DMD area projection | 200/1,000 @ 800 | 14 | 123 | Limited grayscale at the cost of speed | Z axis | 1-4 mJ | Amplifier ~$250,000 |
| Multi-focus[17] | DMD area projection + MLA | 200/600 *3 @ 517 | 1 | 158 | 121 grayscale level | No | 100 nJ | Amplifier ~$250,000 |
| Multi-focus[18] | DMD area projection + DOE + AOD | 163/261 @ 517 | 70 | 123 | No | Z axis | 150 nJ | High-power oscillator ~$150,000 |
| Multi-focus[15] | DMD CGH | 500/1,600 @ 800 | 497 | 0.15 | Limited by algorithm | No | 50 nJ | Oscillator ~$120,000 |
| Multi-focus[16] | DMD CGH | 90/141 @ 800 | 15 | $2.5 \times 10^{-5}$ | Limited by algorithm | No | 4 mJ | Amplifier ~$250,000 |
| Multi-focus[11] | SLM CGH+ Galvo | 215/686 @ 1,030 | 18 | $8.8 \times 10^{-3}$ | Limited by algorithm | No | 10 $\mu$J | Amplifier ~$250,000 |
| Multi-focus[10] | DOE + MLA | 350 *4 @ 790 | 5 | 292 | No | No | 50 nJ | Oscillator ~$120,000 |
| Multi-focus[14] | Metalens + stage scan | 100/160 @ 780 | 1.2 | 100 | No | No | 50 nJ | Oscillator ~$120,000 |

MLA: Micro-lens array

*1: Sustainable printing bandwidth calculation:

The FOV switching time was calculated based on the following parameters: a single FOV size of 0.1 mm, an acceleration of 10 mm/s², and a maximum speed of 1 mm/s. The motion process includes an acceleration phase (0.1 s, with a displacement of 0.05 mm) and a symmetric deceleration phase (0.1 s, with a displacement of 0.05 mm), and there is no uniform motion phase since the total required displacement (0.1 mm) is exactly covered by the acceleration and deceleration stages. Thus, the total motion time reaches 0.2 s. In addition, the overall FOV switching time needs to incorporate the settling time, taking a typical value of 0.05 s as an example, the total FOV switching time is approximately 0.25 s. Additionally, if the lateral FOV movement is performed after z-scan printing, the total FOV switching time can be allocated across 100 layers, which reduces the effective switching time per layer to 0.0025 s.

When a laser amplifier is employed in area projection system, the exposure time is typically set to 0.001 s, which matches the amplifier's repetition rate. For laser oscillators, the exposure time is usually configured as the practical exposure time reported in related studies.

For the CGH multi-focus system, the following parameters were assumed for a single FOV: a lateral scanning resolution of 0.25 μm, a FOV size of 100 μm × 100 μm, a porous structure with 80% porosity. The foci number is adopted as the number reported in related studies, e.g., 4 in Ref[15], 2,000 in Ref[16], and 400 in Ref[11]. For galvo scanning system in Ref[10], the galvo scanner operates at 1 m/s with a 350 nm step and is used in conjunction with the acousto-optic modulator (AOM) driven by a 16-bit (2-byte) digital-to-analog converter (DAC). For stage scanning system in Ref[14], XYZ stages adopt 4-byte floating-point encoding, while the piezo stage has a 10 ms/point positioning latency.

*2: Data preparing rate calculation:

We established pattern generation methods for each study and computed their generation rates. All computations were conducted using MATLAB R2024a (64-bit) on a Windows 11 Pro workstation (Dell Precision 3680), equipped with an Intel Core i5-14600K processor (14 physical cores, 20 logical threads, base clock 3.5 GHz) and 64 GB of memory.

*3: Axial data is not available in the study, we assumed 3 times its lateral size of nanopillar.

*4: Average of lateral and axial size in the study.